\documentclass[]{elsarticle}

\usepackage{amssymb, amsmath, amsthm}
\usepackage{subfigure}
\usepackage{url}

\newtheorem{proposition}{Proposition}

\journal{Some journal}

\begin{document}

\begin{frontmatter}

\title{Excitable London: Street map analysis with Oregonator model}

\author[label1]{Andrew Adamatzky\corref{cor1}}
\address[label1]{Unconventional Computing Lab, Department of Computer Science and Creative Technologies, UWE, Bristol, UK}

\cortext[cor1]{I am corresponding author}
\ead{andrew.adamatzky@uwe.ac.uk}

\author[label1]{Neil Phillips}
\ead{neil.phillips@uwe.ac.uk}

\author[label1,label2]{Roshan Weerasekera}\ead{Roshan.Weerasekera@uwe.ac.uk}
\address[label2]{Department of Engineering, Design and Mathematics, UWE, Bristol, UK}

\author[label1]{Michail-Antisthenis Tsompanas}
\ead{antisthenis.tsompanas@uwe.ac.uk}

\author[label1,label3]{Georgios Ch. Sirakoulis}
\address[label3]{Department of Electrical and Computer Engineering, Democritus University of Thrace, Xanthi, Greece}
\ead{gsirak@ee.duth.gr}

\begin{abstract}
 We explore geometry of London's streets using computational mode of an excitable chemical system, Belousov-Zhabotinsky (BZ) medium. We virtually fill in the streets with a BZ medium and study propagation of excitation waves for a range of excitability parameters, gradual transition from excitable to sub-excitable to non-excitable. We demonstrate a pruning strategy adopted by the medium with decreasing excitability when wider and ballistically appropriate streets are selected. We explain mechanics of streets selection and pruning. The results of the paper will be used in future studies of studying dynamics of cities 
 with living excitable substrates. 
\end{abstract}

\end{frontmatter}


\section{Introduction}
\label{introduction}

Cities are often seen as living organisms in terms of their space-time dynamics of vehicular and pedestrian traffic: cities grow, breath and pulsate as giant creatures~\cite{bettencourt2007growth, hinchliffe2006living, bettencourt2010unified, batty2009centenary, rees2012cities, flynn2016experimenting}. Social and physical processes --- waves of traffic jams~\cite{long2011urban,laval2010mechanism,nagatani2000density,nagatani1999jamming,li2006analysis,jiang2017spatio,sun2018linear,sun2018stability}, and excitation, contagion and diffusion of riots~\cite{bohstedt1994dynamics,myers2000diffusion} --- emerging on city streets often resemble, and not rarely shared biological and physical mechanisms with excitation waves in chemical media~\cite{tyson1988singular,jung1998noise} or cultures/tissues of living cells~\cite{steinbock1993three,holden2013nonlinear,mikhailov1991kinematical,seiden2015tongue}. To study spatial and temporal dynamics of these wave phenomena, we  consider that city streets are filled with excitable chemical system --- the Belousov-Zhabotinsky medium~\cite{belousov1959periodic, zhabotinsky1964periodic}.

A thin-layer BZ medium shows rich dynamics of excitation waves, including target waves, spiral waves and localised wave-fragments and their combinations. A substantial number of theoretical and experimental laboratory prototypes of computing devices made of BZ medium has been reported in the last thirty years. Most interesting include logical gates implemented in geometrically constrained BZ medium~\cite{steinbock1996chemical, sielewiesiuk2001logical}, memory in BZ micro-emulsion \cite{kaminaga2006reaction}, information coding with frequency of oscillations~\cite{gorecki2014information},  chemical diodes~\cite{DBLP:journals/ijuc/IgarashiG11}, neuromorphic architectures~\cite{ gorecki2006information, gentili2012belousov, takigawa2011dendritic, stovold2012simulating,  gruenert2015understanding}, associative memory~\cite{stovold2016reaction,stovold2017associative}, wave-based counters~\cite{gorecki2003chemical}, evolving logical gates~\cite{toth2009experimental} and binary arithmetic circuits~\cite{costello2011towards, sun2013multi, zhang2012towards, suncrossover, digitalcomparator}. By controlling BZ medium excitablity we can produce  related  analogs of dendritic trees~\cite{takigawa2011dendritic}, polymorphic logical gates~\cite{adamatzky2011polymorphic} and logical circuits~\cite{stevens2012time}. Light sensitive modification, with  Ru(bpy)$^{\text{+3}}_{\text 2}$ as catalyst, allows for manipulation of the medium excitablity, and subsequent modification of geometry of excitation wave fronts~\cite{kuhnert1986new, braune1993compound, manz2002excitation}. A light-sensitive BZ medium allows for optical inputs of information as parallel inputs in massive parallel processors. The medium can be also constrained geometrically in networks of conductive channels, thus allowing for a directed routing of signals. This is why we will be using  two-variable Oregonator model~\cite{field1974oscillations} adapted to a light-sensitive Belousov-Zhabotinsky (BZ) reaction with applied illumination~\cite{beato2003pulse}. Approximation of shortest path, also in the context of maze solving problem, with BZ was studied experimentally in~\cite{agladze1997finding, steinbock1995navigating, rambidi1999finding}, approximation of distance fields used in robot navigation \cite{adamatzky2002collision}, and partly applied in 
on-board controllers for robots~\cite{adamatzky2004experimental, DBLP:journals/ijuc/Vazquez-OteroFDD14}. All previous theoretical and experimental works on space-exploration with BZ medium dealt with the medium in excitable model. When exploring a geometrically constrained space filled with excitable BZ medium excitation wave fronts typically propagate in all directions, exploring all possible channels, as a flood fills.  Excitation waves in sub-excitable medium behave sometimes as localised patterns, quasi-dissipative solitons, and, thus, they might not explore all space available~\cite{hildebrand2001spatial, adamatzky2004collision, de2009implementation}. 

The paper is structured as follows. We introduce Oregonator model of a light-sensitive BZ reaction and other components of modelling in Sect.~\ref{methods}. We exemplify and discuss spatial dynamics of excitation on London streets template in Sect.~\ref{dynamics}. We uncover mechanisms of streets spanning by wave of excitation and analyses ballistic properties of wave-fragments in sub-excitable BZ medium in Sect.~\ref{mechanics}. We discuss outcomes of the studies in Sect.~\ref{discussion}.

\section{Methods}
\label{methods}

\begin{figure}[!tbp]
\centering
\includegraphics[]{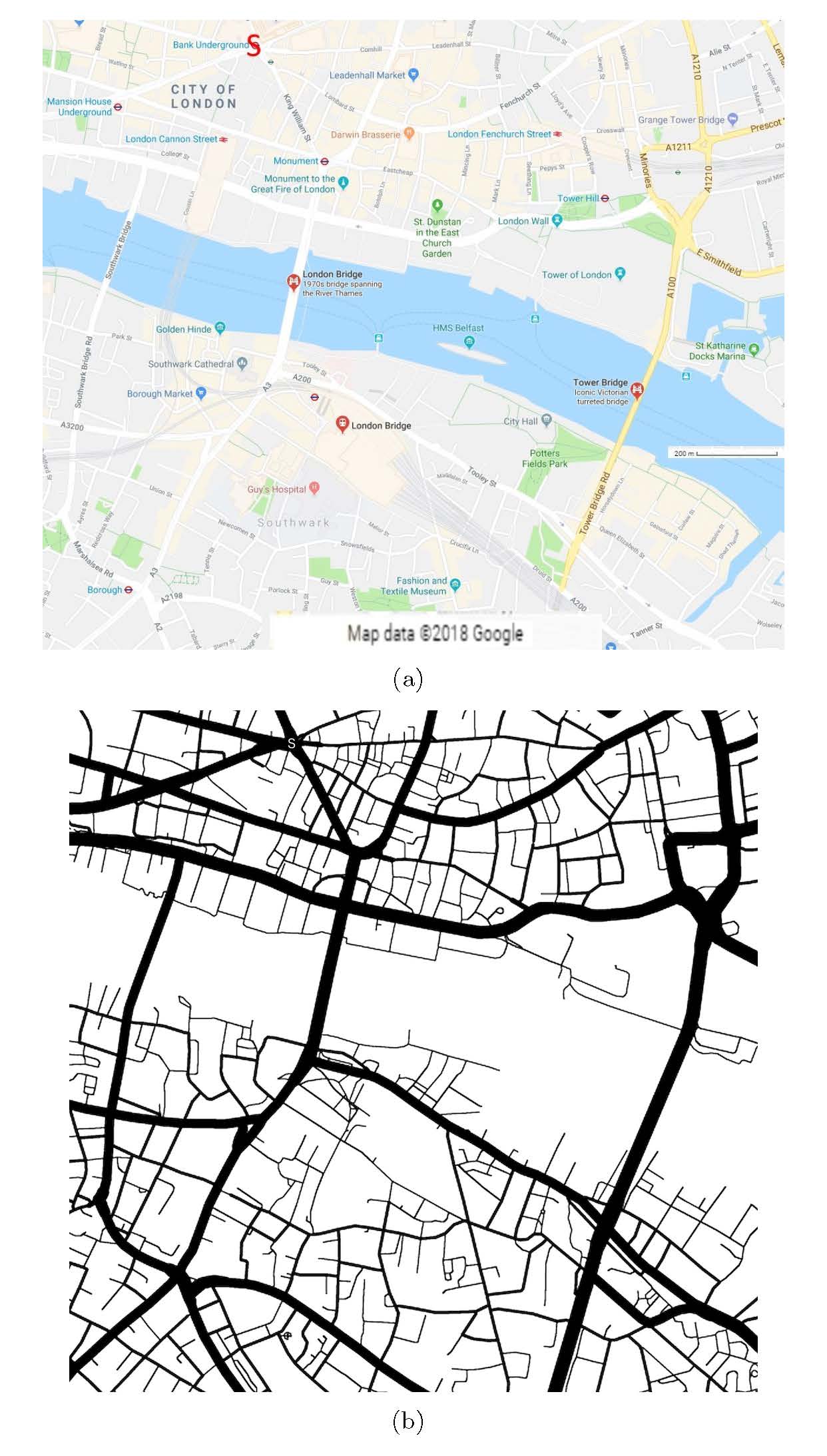}
\caption{Fragment of London street map used in computational experiments. Position of initial excitation is shown by `S' in the top of the North-West quadrant. 
(a)~Google map \protect\url{https://www.google.co.uk/maps/search/london+bridge/@51.5072749,-0.0906453,15.75z} \copyright 2018 Google. (b)~Template used for studies.}
\label{map}
\end{figure}

A fragment of London street map (map data \copyright 2018 Google) approximately 2 km by 2 km, centred around London Bridge, was mapped onto a grid of 2205 by 2183 nodes (Fig.~\ref{map}). Nodes of the grid corresponding to streets are considered to be filled with Belousov-Zhabotinsky medium, i.e. excitable nodes, other nodes are non-excitable (Dirichlet boundary conditions, where value of variables are fixed zero). We use two-variable Oregonator equations~\cite{field1974oscillations} adapted to a light-sensitive 
Belousov-Zhabotinsky (BZ) reaction with applied illumination~\cite{beato2003pulse}:
\begin{eqnarray}
  \frac{\partial u}{\partial t} & = & \frac{1}{\epsilon} (u - u^2 - (f v + \phi)\frac{u-q}{u+q}) + D_u \nabla^2 u \nonumber \\
  \frac{\partial v}{\partial t} & = & u - v 
\label{equ:oregonator}
\end{eqnarray}

The variables $u$ and $v$ represent local concentrations of an activator, or an excitatory component of BZ system, and an inhibitor, or a refractory component. Parameter $\epsilon$ sets up a ratio of time scale of variables $u$ and $v$, $q$ is a scaling parameter depending on rates of activation/propagation and inhibition, $f$ is a stoichiometric coefficient. 
Constant $\phi$ is a rate of inhibitor production.  In a light-sensitive BZ $\phi$ represents the rate of inhibitor production proportional to intensity of illumination. The parameter $\phi$ characterises excitability of the simulated medium. The larger $\phi$ the less excitable medium is. We integrated the system using Euler method with five-node Laplace operator, time step $\Delta t=0.001$ and grid point spacing $\Delta x = 0.25$, $\epsilon=0.02$, $f=1.4$, $q=0.002$. We varied value of $\phi$ from the interval $\Phi=[0.05,0.08]$. The model has been verified by us in experimental laboratory studies of BZ system, and  the sufficiently satisfactory match between the model and the experiments was demonstrated in \cite{adamatzky2007binary, de2009implementation, toth2010simple, adamatzky2011towards}. 
To generate excitation wave-fragments we perturb the medium by a square solid domains of excitation, $20 \times 20$ sites in state $u=1.0$. Time lapse snapshots provided in the paper were recorded at every 150\textsuperscript{th} time step, we display sites with $u >0.04$; videos supplementing figures were produced by saving a frame of the simulation every 50\textsuperscript{th} step of numerical integration and assembling them in the video with play  rate 30 fps.  All figures in this paper show time lapsed snapshots of waves, initiated just once from a single source of stimulation; these are not trains of waves following each other.

\section{Dynamics of excitation}
\label{dynamics}

\begin{figure}[!tbp]
    \centering
\includegraphics[]{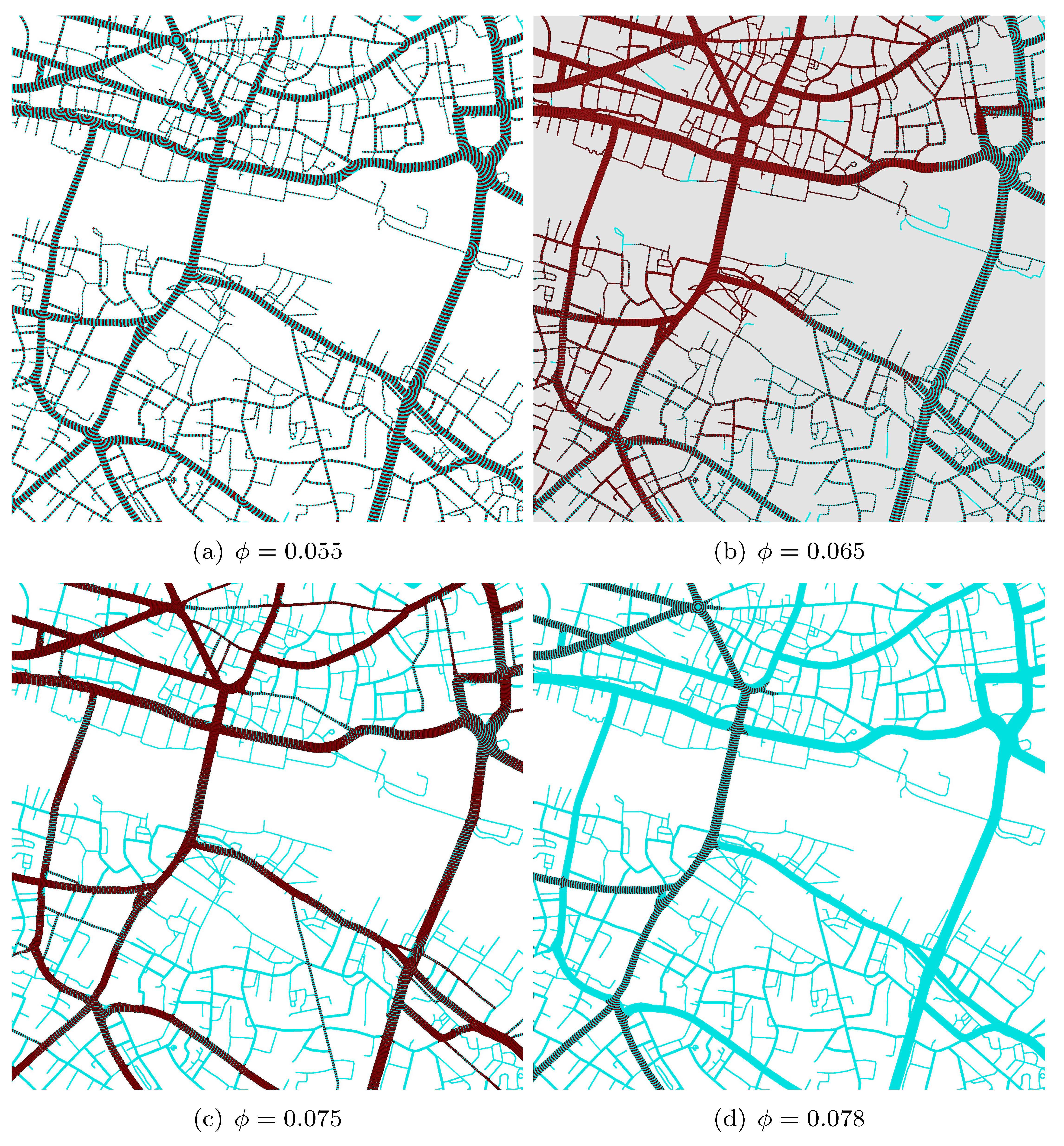}
    \caption{Propagation of excitation on the street map. These are time lapsed snapshots of a single wave-fragment recorded every 150\textsuperscript{th} step of numerical integration. Values of $\phi$ are shown in captions. }
    \label{fig:lapseexamples}
\end{figure}

\begin{figure}[!tbp]
    \centering
    \includegraphics[]{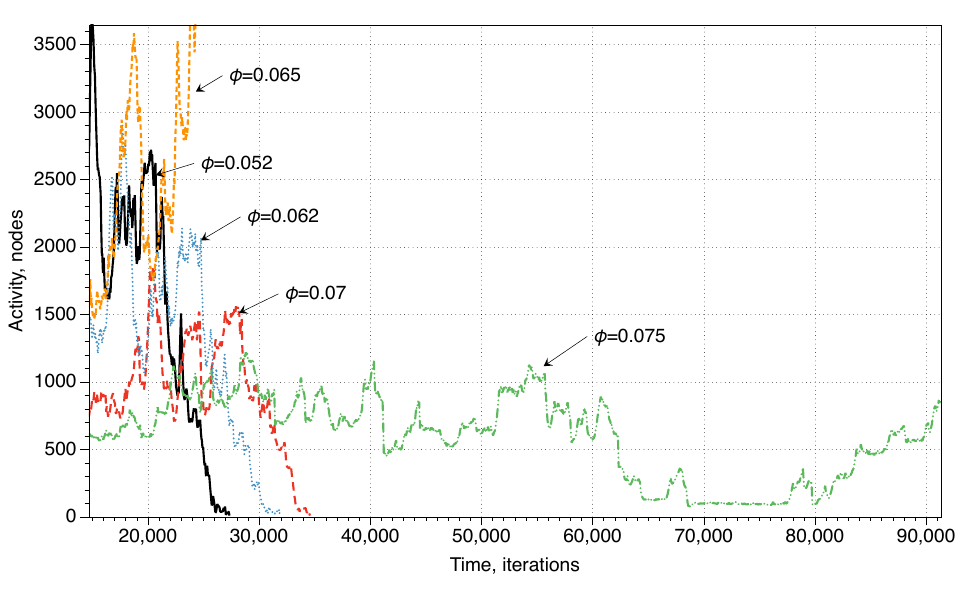}
    \caption{Integral dynamics of excitation calculated as a number of grid nodes with $u>0.1$ at each time step of integration. The dynamics is shown for the media with $\phi=$0.52, 0.62, 0.65, 0.7, 0.75. The dynamics is show till the moments when excitation waves disappear or all streets got covered, whichever occurs earlier.}
     \label{fig:integralexcitation}
     \end{figure}

\begin{figure}[!tbp]
    \centering
    \includegraphics[]{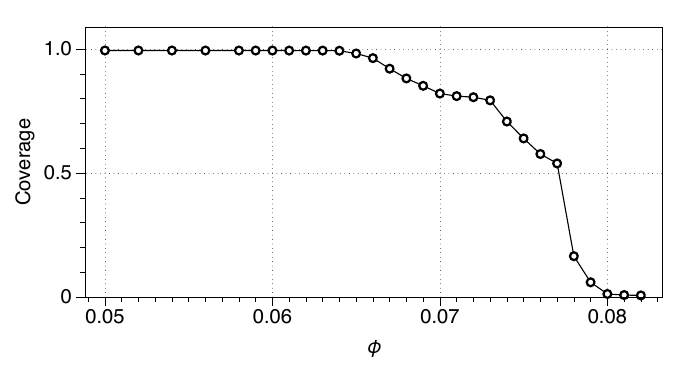}
\caption{Coverage of the street network by excitation waves for $\phi \in \Phi$. A value of coverage is calculated as a ratio of nodes, representing streets, excited ($u>0.1$) at least once during the medium's evolution to a total number of nodes representing streets. Values of coverage are displayed by circles for $\phi$ increasing from 0.05 with increment 0.001.}
    \label{fig:coverage_vs_phi}
\end{figure}

Excitable medium with values of $\phi$ from lower end of interval $\Phi$ exhibit excitation waves propagating along all streets, independently on their width, and passes junction where streets join each at various angles (Fig.~\ref{fig:lapseexamples}a). The excitation disappears when all waves reach boundaries of the map.  Integral dynamics of excitation is characterised by a sharp increase in a number of excited nodes followed by abrupt disappearance of the excitation when all wave fronts leave the lattice (e.g. Fig.~\ref{fig:integralexcitation}, $\phi=0.052$). 

With increase of $\phi$ to the middle of $\Phi$ we observe repeated propagation of excitation wave fronts along some parts of the map. Paths of the streets where excitation cycles are seen as having higher density of wave-fronts in (Fig.~\ref{fig:lapseexamples}b). Two peaks in the excitation activity seen in the plot $\phi=0.065$ (Fig.~\ref{fig:integralexcitation}) are due to a combination of two factors: structure of the street network and cycles of excitation emerged. The street map graph consists of Northern and Southern parts connected by few bridges over Thames. The medium is excited in its northernmost part domain of the street map (Fig.~\ref{map}). Therefore, by the time excitation wave fronts cross bridges into the Southern part the excitation propagates along all streets in the Northern part and disappear across the borders of the integration grid. When excitation in the Northern part disappears, just after 20,000\textsuperscript{th} step of integration, we observe sudden drop in the number of excited nodes. The streets in the Southern part are then getting excited.

Value $\phi=0.064$ is the highest for which all streets are covered by excitation wave fronts (Fig.~\ref{fig:coverage_vs_phi}). For $\phi=0.075$ excitation waves propagate only along major streets (Fig.~\ref{fig:lapseexamples}c). And just few of the major streets are selected by excitation for $\phi=0.078$ (Fig.~\ref{fig:lapseexamples}d). Due to excitation following only selected paths an overall time the medium stays excited becomes substantial, up to 100k of iterations (Fig.~\ref{fig:integralexcitation}).

\begin{figure}[!tbp]
    \centering
    \includegraphics[]{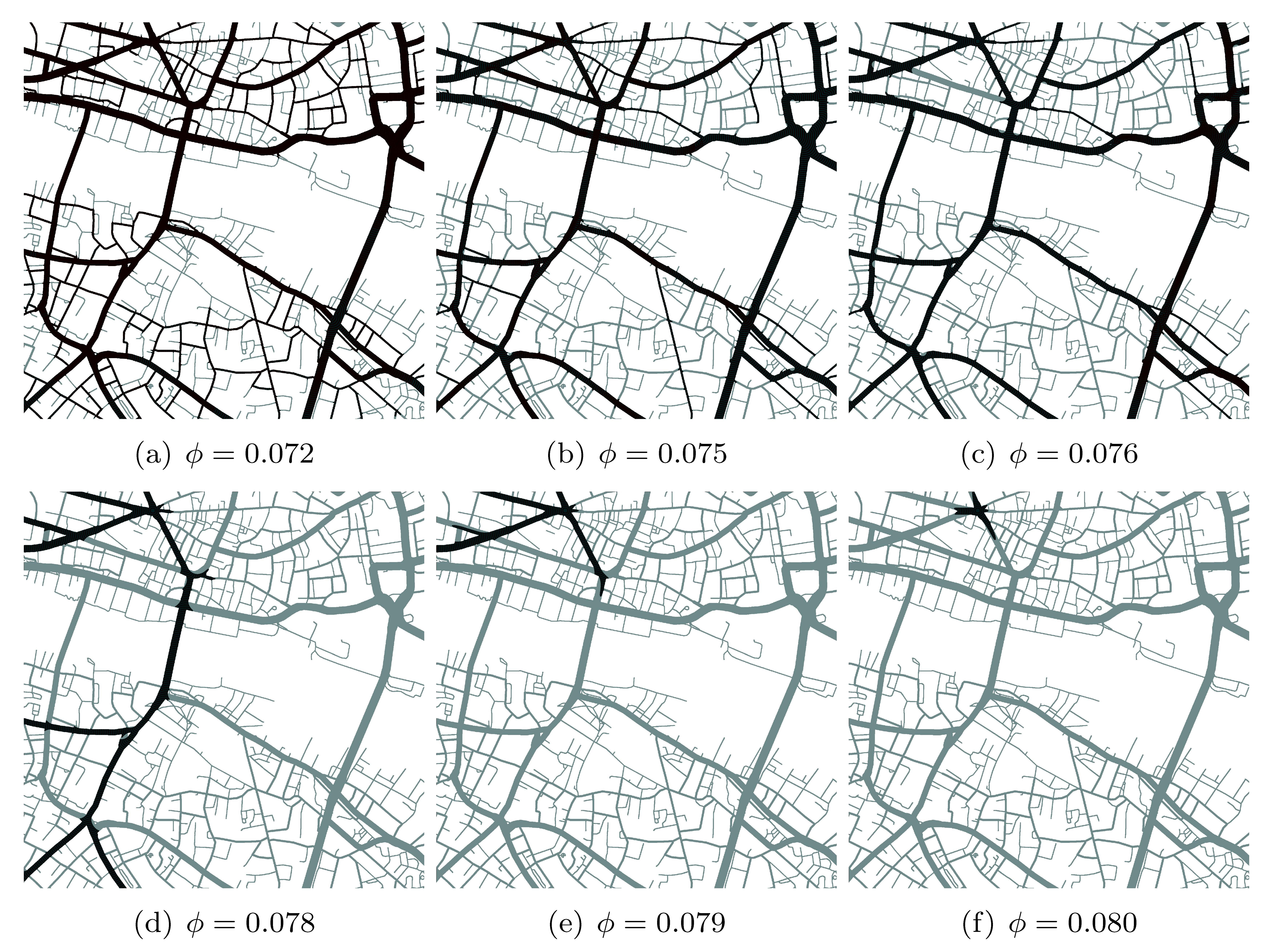}
    \caption{Spatial coverage of street networks by excitation wave fronts for $\phi$ from 0.072 to 0.080. Nodes, representing streets, which have been excited at least once during the medium evolution are black. Nodes which never been excited are light grey.
    }
    \label{fig:pruning}
\end{figure}

Spatial coverage of street networks by excitation wave fronts for various values of $\phi$ is decreasing with the decrease of the medium's excitability (increase of $\phi$), as illustrated in Fig.~\ref{fig:pruning}. Looking at Fig.~\ref{fig:pruning} one might think that by decreasing excitability we prune the street network by selecting only widest streets. This is partly but not always true. The explanations are in the next section.

\section{Mechanics of exploration}
\label{mechanics}

\begin{figure}[!tbp]
\centering
\includegraphics[]{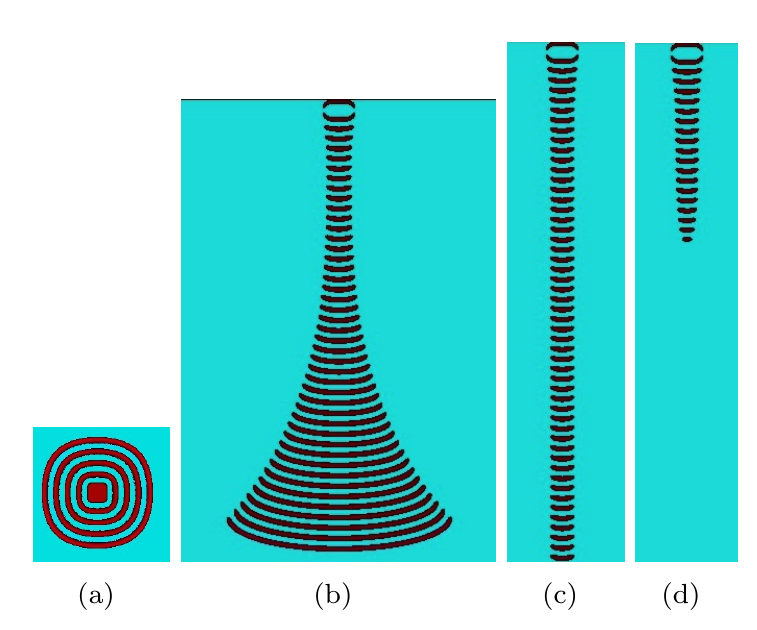}
\caption{Types of waves. All site of the medium are conducive for excitation.
(a)~Classical wave in an excitable medium, $\phi=0.05$.
(bcd)~Behaviour of an initially asymmetric excitation in BZ medium for different levels of excitability.
(b)~Expanding wave-fragment, $\phi=0.076690$,
(c)~Shape preserving wave-fragment, $\phi=0.076691$,
(d)~Collapsing wave-fragment, $\phi=0.076699$. These are time lapsed snapshots of a single wave-fragment recorded every 150\textsuperscript{th} step of numerical integration.}
\label{waves}
\end{figure}

\begin{figure}[!tbp]
    \centering
    \includegraphics[]{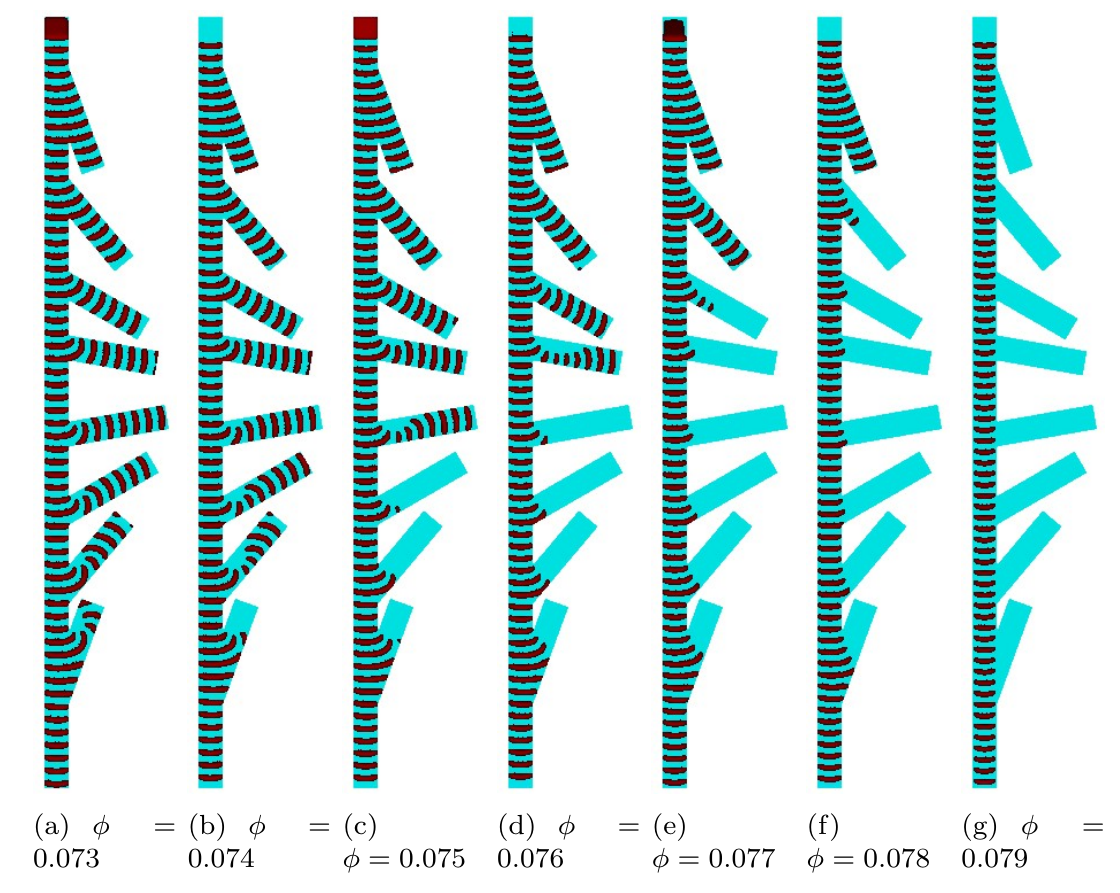}
    \caption{Excitation waves enter side streets depending on the medium's excitability. Excitation is initiated at the North top of the vertical channel. Side channels join the vertical channel at angles, 20\textsuperscript{o} to 160\textsuperscript{o} with increement 20\textsuperscript{o}.  Values of $\phi$ are shown in captions. }
    \label{fig:angles}
\end{figure}

\

Dynamical regimes of BZ medium can be subdivided to excitable $\phi<0.064$, 
sub-excitable $0.064 \leq \phi < 0.08$  and non-excitable  $0.08 \leq \phi$. In excitable mode a perturbation leads to formation of circular wave fronts (Fig.~\ref{waves}a) which propagates in all directions.  If streets are filled with excitable medium then, independently on a site of initial perturbation excitation travels to all streets.  In a sub-excitable medium we can observe three regimes of the medium response to asymmetric perturbation: expanding wave fronts, $0.0766 \leq \phi \leq 0.076690$ (Fig.~\ref{waves}b), shape preserving wave fronts,$0.076691 \leq \phi \leq 0.076698$  (Fig.~\ref{waves}c) and collapsing wave fronts, $0.076691 \leq \phi \leq 0.076699$  (Fig.~\ref{waves}d). 

\begin{proposition}
A coverage of a street network by an excitation originated in an arbitrary point of the street network filled with BZ medium with $\phi \geq 0.076$ is proportional to a ratio of junctions with street branching at acute angles met by propagating wave fronts. 
\end{proposition}

\begin{figure}
    \centering
    \includegraphics[]{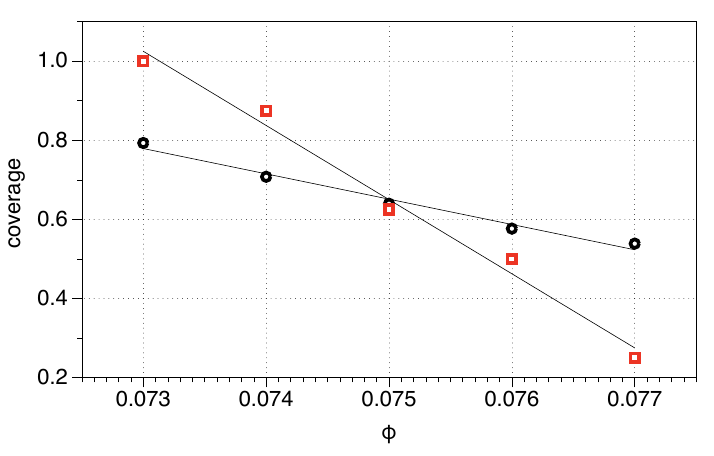}
    \caption{Coverage of London street network, Fig.\ref{map}, black circles, and structure with equal number of acute and obtuse junctions, Fig.~\ref{fig:angles}, red squares, by excitation waves in BZ medium with $\phi=0.073$ to 0.077. Lines are linear fits.}
    \label{fig:comparecoverage}
\end{figure}

Formal proof of this proposition will be scope of future studies. Here we consider constructive support of the statement. All three types of sub-excitable waves propagate alike at the initial stages of their evolution: they preserve their shape and velocity vector. When such waves collide into non-excitable object they do not reflect but partly annihilate. If --- depending on an angle of collision --- some part of the wave front did not come in contact with non-excitable object it might restore its shape, especially if the medium is in the expanding wave mode, and continue its propagation. Shape preservation and ballistic propagation of wave-fragments in sub-excitable medium for higher values of $\phi$ ($\phi \geq 0.076$) prevent excitation from entering site streets. Chances that a wave-front enters a side street are proportional to angle between wave-front velocity vector and a vector from the main street to the side street. This is illustrated in Fig.~\ref{fig:angles}.  Coverage versus $\phi$ plots for street network Fig.\ref{map} and test structure Fig.~\ref{fig:angles} are shown in Fig.~\ref{fig:comparecoverage}. Linear fits of the plots are as follow: 
coverage\textsubscript{street}~=~$-63.994\cdot\phi$ + 5.4507 and 
coverage\textsubscript{test}~=~$-187.5\cdot\phi$ + 14.713. The test structure Fig.~\ref{fig:angles} has equal numbers of junctions with acute and obtuse angles. Absolute value of the slop of coverage\textsubscript{street} is lower than that of the slop of coverage\textsubscript{test}. This might indicate that an excitation wave fronts propagating from the chosen site of perturbation (labelled by 's' in Fig.\ref{map}) encounter more junctions with acute angles. Indeed, we could expect different slop of a coverage for another site of initial perturbation, that would be a matter of further studies.

\begin{figure}[!tbp]
\includegraphics[]{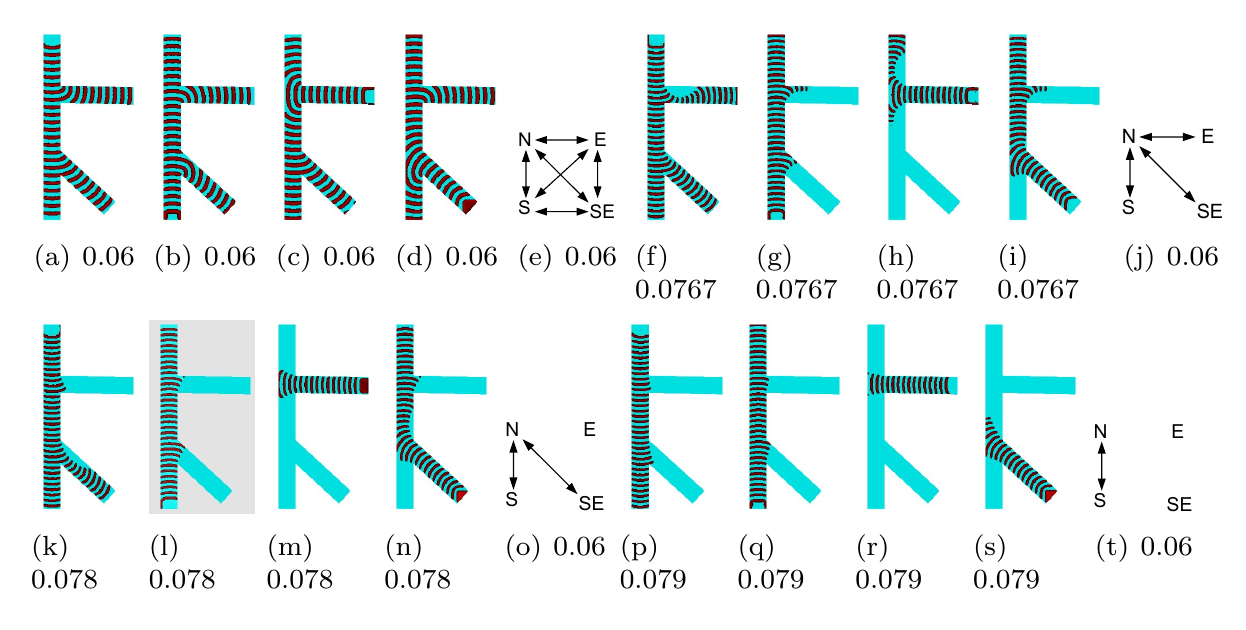}
    \caption{Time lapse snapshots of excitation waves initiated at end of North, `N', segment (a,f,k,p); South, `S', segment (b,g,l,q); East, `E', (c,h,m,r); South-East, `SE' (d,i,n,s). Values of $\phi$ are shown in the captions. These are time lapsed snapshots of a single wave-fragment recorded every 150\textsuperscript{th} step of numerical integration.}
    \label{fig:commutativity}
\end{figure}

A network of channels filled with BZ medium is commutative if the following condition takes place: if excitation wave front originated by a perturbation of site $a$ reaches site $b$ then excitation wave front originated by a perturbation of site $b$ reaches site $a$. If BZ  medium is in its excitable mode then the network is commutative. To check if the same holds for sub-excitable BZ medium we constructed a template of three channels shown in Fig.~\ref{fig:commutativity}. We found that the template is commutative. A graph of reachability though is pruned with increase of $\phi$. The graph (Fig.~\ref{fig:commutativity}e) is fully connected in excitable medium, $\phi=0.06$ (Fig.~\ref{fig:commutativity}a--d), tree (Fig.~\ref{fig:commutativity}j) is obtained for $\phi=0.0767$ (Fig.~\ref{fig:commutativity}f--i), chain and one isolated node (Fig.~\ref{fig:commutativity}o) for $\phi=0.078$ (Fig.~\ref{fig:commutativity}k--n), a single edge and two isolated nodes  (Fig.~\ref{fig:commutativity}t) is the result for $\phi=0.079$ (Fig.~\ref{fig:commutativity}p--s). There is always a chance that we missed some configuration of channels which demonstrates non-commutativity for some values $\phi$.

\section{Discussion}
\label{discussion}

\begin{figure}[!tbp]
    \centering
    \includegraphics[]{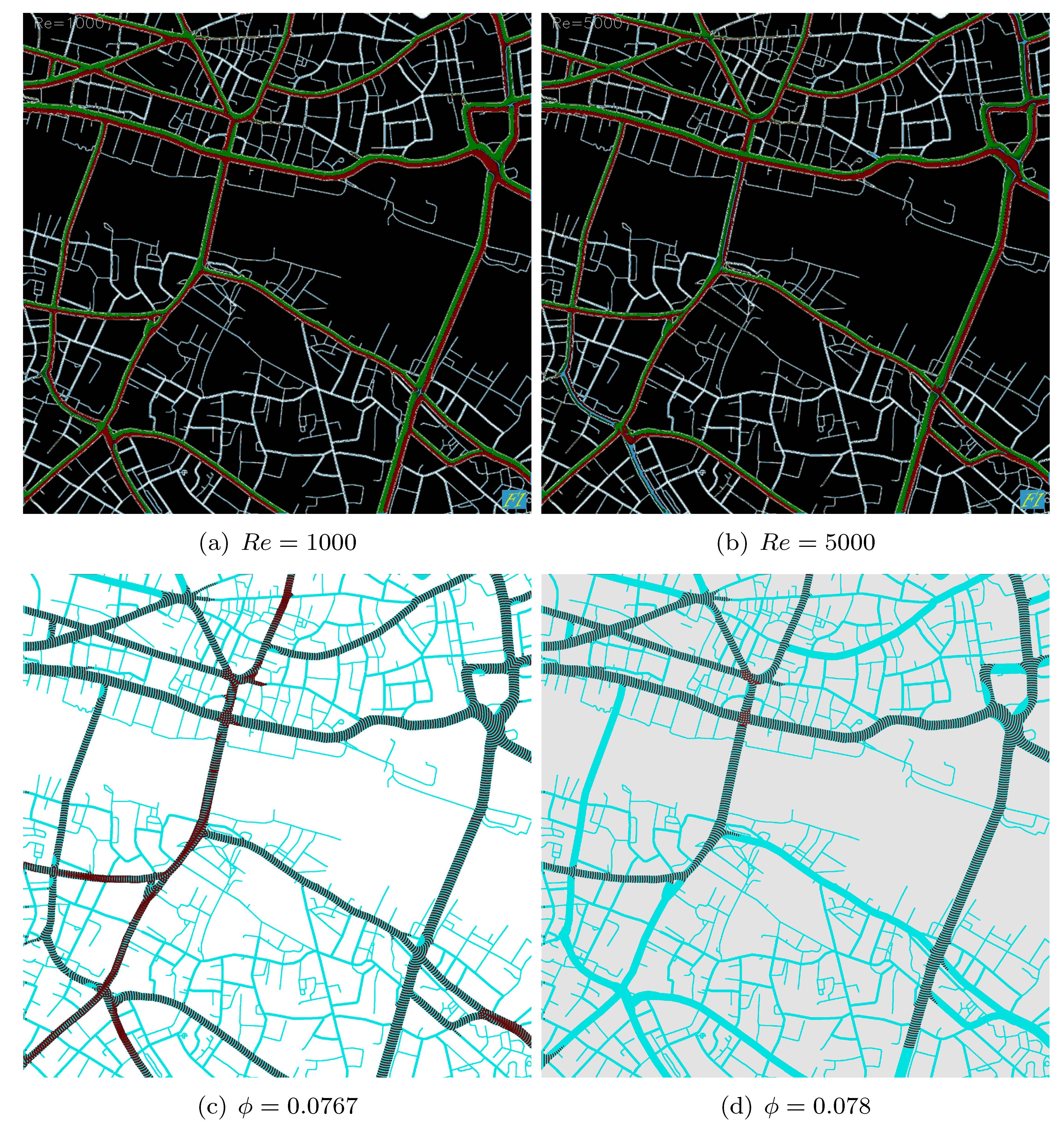}
    \caption{Comparison of (ab) laminar fluid flow for (a) low Reynolds number $Re=1000$ and (b) high Reynolds number $Re=5000$  with excitation propagation for (c) $\phi=0.0767$ and (d) $\phi=0.078$. Fluid stream and excitation enter street networks at streets on the Northern edge.
    (ab) are  snapshots of fluid streaming along streets produced in Fluid Flow Illustrator (\protect\url{http://www.flowillustrator.com/}).
    (cd) are time lapsed snapshots of a wave-fragments recorded every 150\textsuperscript{th} step of numerical integration.
    }
    \label{fig:fluidflow}
\end{figure}

Using numerical integration of Oregonator model of Belousov-Zhabotinsky medium in a fragment of London street network we demonstrated that 
(1)~coverage of the network is proportional to excitability of the medium, 
(2)~cycling patterns of excitation are only possible in sub-excitable regime of the medium, 
(3)~wave-fragments in sub-excitable network propagate ballistically, 
(4)~coverage of the street network by excitation in a medium with a given value $\phi$ is proportional to a ratio of junctions with streets branching out at acute angle to junctions with streets branching out at obtuse angles, 
(5)~reachability by excitation wave-fronts is commutative.

Excitation wave fragments propagate ballistically. Close analogies could be a crowd charging along the streets or fluid jet streams propagating along the streets. To evaluate the analogy with a fluid flow we simulate jet streams entering streets from the Western edge of the street map domain and leaving in all other edges. We compared the dynamics of fluid flow with dynamics of excitation waves initiated at the Western edge of the integration grid. There is nearly a perfect match between fluid flow for Reynolds number $Re=1000$ (Fig.~\ref{fig:fluidflow}a) and sub-excitable medium with $\phi=0.0767$ (Fig.~\ref{fig:fluidflow}c), with just one street (top of the north-west quadrant) covered by the a jet stream and not covered by excitation wave.  Increase of Reynolds number leads to the effect of street pruning as occurs with increase of $\phi$ (Fig.~\ref{fig:fluidflow}b, d), however the coverage of streets by excitation waves is substantially different from that by fluid flow: 
the only match in the two streets in the south-west quadrant where flow is disrupted by turbulence (blue coloured in Fig.~\ref{fig:fluidflow}b)  and no excitation is travelling the same streets (Fig.~\ref{fig:fluidflow}d). 
1
Ballistic behaviour of the excitation-wave fragments is somewhat similar to a herding behaviour of crowds observed during various scenarios of evacuation, especially when stress 
is involved \cite{zheng2009modeling,burger2011continuous,helbing2000simulating,vihas}. The herding behaviour might lead to situations, when panic stricken crowd traps itself in potentially dangerous domains of the space~\cite{helbing2002simulation,helbing2002crowd}. Exploring geometries with sub-excitable medium could help us to identify loci of the space when a crowd charging in a herding mode could find itself stuck. Pruning behaviour sub-excitable BZ medium for high values of $\phi$ might also help to assess areas of a geometric network which would more frequently to be visited by living creature: this is based on our previous studies of space exploration by leeches~\cite{adamatzky2015exploration,adamatzky2015building}.

\clearpage

\bibliographystyle{plain}
\bibliography{bibliography}

\end{document}